\documentclass[
superscriptaddress,
reprint,
amsmath,amssymb,
aps,
prd,
]{revtex4-1}

\usepackage[normalem]{ulem}
\usepackage{hyperref,xcolor}
\usepackage{color}

\usepackage{physics}
\usepackage{graphicx}
\usepackage{dcolumn}
\usepackage{bm}
\usepackage{amsthm}
\usepackage{tikz}

\usepackage{bbm}

\newcommand{\id}{{\mathbbm 1}}
\def\d{{\rm d}}

\begin{document}
\title{Quantum harvester enables energy transfer  without randomness transfer or dissipation}

\author{Fei Meng}
\affiliation{Department of Physics, City University of Hong Kong, Tat Chee Avenue, Kowloon, Hong Kong SAR}
\author{Junhao Xu}
\affiliation{Shenzhen Institute for Quantum Science and Engineering and Department of Physics, SUSTech, Nanshan District, Shenzhen, China.}
\author{Xiangjing Liu}
\affiliation{Shenzhen Institute for Quantum Science and Engineering and Department of Physics, SUSTech, Nanshan District, Shenzhen, China.}
\author{Oscar Dahlsten}
\email{oscar.dahlsten@cityu.edu.hk}
\affiliation{Department of Physics, City University of Hong Kong, Tat Chee Avenue, Kowloon, Hong Kong SAR}
\affiliation{Shenzhen Institute for Quantum Science and Engineering and Department of Physics, SUSTech, Nanshan District, Shenzhen, China.}
\affiliation{Institute of Nanoscience and Applications, Southern University of Science and Technology, Shenzhen 518055, China}

\begin{abstract}
We consider a foundational question in energy harvesting: given a partly random energy source, is it possible to extract the energy without also transferring randomness or accepting another thermodynamical cost? We answer this in the positive, describing scenarios and protocols where in principle energy is extracted from a field with randomness but without any randomness being transferred, and without energy dissipation. Such protocols fundamentally outperform existing methods of rectification which dissipate power, or feedback demon-like protocols which transfer randomness to the feedback system. The protocols exploit the possibility of the harvesting system taking several trajectories that lead to the same final state at a given time. We explain why these protocols do not violate basic physical principles. A key example involves the experimentally well-established phenomenon of Rabi oscillations between energy levels, exploiting the multitude of rotation axes in the state space that take the lower energy state to the excited state. The quantum system is deterministically excited to the highest energy level after interacting with the source for a fixed amount of time, irrespective of the random initial phase of the external potential.
\end{abstract}
\maketitle

\textbf{Introduction.} Harvesting energy from the ambient environment is a technologically important task. For example, there is great interest in developing small harvesters to provide autonomous power supply for mobile electronic devices~\cite{Mitcheson2008,wei2017comprehensive}. Typically the harvester converts a force from the ambient environment to useful energy, e.g. a piezo-electric material converting mechanical force to an electrical voltage or an antenna harvesting energy from an oscillating ambient electric field~\cite{Mitcheson2008,wei2017comprehensive}. A key challenge is that real-world sources are stochastic, varying and sometimes transient. To remove randomness in the voltage output, one standard method is to employ rectifiers  $V\rightarrow |V|-\delta$ at some penalty $\delta$~\cite{horowitz1989art,Szarka2012,surender2021rectenna}. The penalty $\delta$ is associated with dissipation in the diodes making up the rectifier, and is fundamentally non-zero in that otherwise a rectifier could extract energy from thermal fluctuations~\cite{liu2019intelligently}. Another method under investigation is to intervene with active and carefully timed voltage flips which requires, unless the signal is predictable,  measuring the voltage~\cite{Cao2007, Siebert2005}, fundamentally transferring randomness to the system implementing the measurement and feedback system~\cite{bennett1982thermodynamics}.

The rapidly developing quantum technology opens new possibilities. Quantum technologies involve the control of quantum superpositions of states, similar to yet distinct from probabilistic combinations of states~\cite{nielsen2010quantum}. Quantum technology can be applied for sensing, computation, and communication~\cite{Acín_2018,degen2017quantum,arute2019quantum}. An area of intense current research concerns the possibility of advantages in energy efficiency--see for example Refs.~\cite{auffeves2022quantum, Jaramillo_2016,chiribella2022nonequilibrium,Shi2022,felce2020quantum,liu2022thermodynamics,Alexia2023experimental,jaliel2019experimental,thierschmann2015three,Alexia2022energetic}.

\begin{figure}     
\centering
\includegraphics[width=\linewidth]{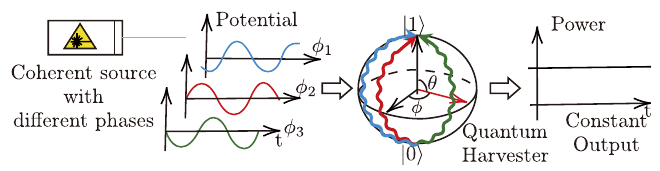}
    \caption{\textbf{A two-level quantum system removes randomness in the sinusoidal source phase.} In a suitable basis, the evolution at resonance is rotation along an axis on the equator ($\theta = \pi /2$), and thus the ground state $\ket{0}$ is rotated to the excited state $\ket{1}$ after a fixed amount of time, irrespective of $\phi$. Then a unitary can be applied to restore $\ket{1}$ back to $\ket{0}$ while outputting constant power.}
\label{fig:energy_harvesting_two_phases}
\end{figure}

We here show that quantum technology can be used as an interface that removes randomness in the potential from the ambient environment. As a paradigmatic example, we consider an oscillating potential with unknown shift $\phi$ as in Fig.~\ref{fig:energy_harvesting_two_phases}. By suitable choice of a quantum harvesting system, the quantum system can then in principle transition from a lower energy level to a higher energy level deterministically at a fixed later time. Different phases $\phi$ are associated with different trajectories that all lead (up to a global phase) to the same final state. We generalise the paradigmatic example in terms of quantum system dimension and potential profiles. We describe how such energy transfer without randomness transfer is consistent with the second law and Liouville's theorem. We give a quantum-inspired classical mechanical protocol that also in principle achieves energy transfer without randomness transfer.

The setting of sinusoidally varying potential is realistic and the quantum control required is within the established capabilities of current quantum technology experiments.

\textbf{Energy harvesting protocol.}  
Consider a random energy source interacting with a harvester with generalised position $q$ modelled by a set of time-(in)dependent potentials $\mathcal{V}=\{V_{\phi}(q,t)\}$ with each potential labelled by a real number $\phi$. Each time when we try to harvest energy, $V_{\phi}(q,t)$ is given as a black box with probability $\Pr(\phi)$. 

We assume that the potential source is stable enough not to be affected by the back-action of a system interacting with it. A particularly important example, widely existing in nature and in energy harvesting models is a sinusoidal potential $V_{\phi}(t) = V_0 \cos (\omega t + \phi)$, which can be used to model magnetic, electric, and mechanical oscillations and AC voltage sources with random initial phases. The total Hamiltonian of the harvester system is 
\begin{equation}
H_{tot}= H_0 + V_{\phi}(t) \, .
\end{equation}
where $H_0$ can be termed the bare Hamiltonian, independently of the external force. 

We shall focus on cyclical protocols, which are divided into two stages. The first stage changes the system state $\rho_0\rightarrow\rho_T$ in an approximately deterministic manner but possibly via many trajectories associated with different external potentials $V_{\phi}$. The second stage is the deterministic lowering of energy associated with $\rho_T\rightarrow\rho_0$ due to interaction with the load. {Our analysis is focussed on removing the randomness associated with the source in the first stage}. The harvester starts with a state whose energy is $E_0$ and ends up with energy $E_T$, such that the amount of energy harvested is $\Delta E= E_T - E_0$.

\medskip
\textbf{Deterministic energy harvesting (DEH)}.
We call a protocol {\em deterministic energy harvesting} (DEH) with respect to the set of random source $\mathcal{V}$ parametrized by a random variable $\phi$ if the protocol has the following properties: (i) Open-loop: the protocol does not use active feedback control. (ii) Fixed evolution time: the time spent $T$ to implement the protocol is the same for all random potentials.
(iii) Fixed state transition: the protocol starts with a fixed initial state $\rho_0$ and reaches the same final state $\rho_T$ irrespective of which random potential is present.  The conditions (i) and (iii) combined imply that the amount of energy $\Delta E$ transferred to the load does not depend on the random variable $\phi$ and that there is no need to actively control the protocol to harvest a deterministic amount of energy. Accordingly,  for continuous dependence on $\phi$, DEH requires that
\begin{equation}
\label{eq:DEH}
    \frac{\partial \Delta E}{\partial \phi} =\frac{\partial \rho_0}{\partial \phi}=\frac{\partial \rho_T}{\partial \phi} = 0 \, .
\end{equation}
For example, DEH protocols exist for $\mathcal{V}=\{ V(q,t)+f(\phi)|\phi \in \mathbb{R} \}$, because the $f(\phi)$ induces a constant shift to the Hamiltonian and has no physical effect---there is one and only one state trajectory. 

\medskip
\textbf{Standard oscillator harvesters fail to achieve DEH for important case.}
Typical energy harvesters give an output that depends on the oscillating potential's phase shift $\phi$. Consider the case of a potential $V(t) = - F_0 \cos (\omega t + \phi ) q$, where $F_0$ is the maximum force. A standard approach to harvesting energy from such a potential is to use a linear oscillator with bare Hamiltonian $H_0=p^2/2m + kq^2/2$,  for mass $m$, spring constant $k$ and $\sqrt{k/m}=\omega$ such that there is resonance~\cite{yeatman2009energy}. By Newton's second law, $m \Ddot{q}=-k q+F_0 \cos (\omega t + \phi)$. Assuming initially for simplicity that $q(0)=0$ and velocity $\dot{q}(0)=0$, solving the ordinary differential equation gives
\begin{equation}
\label{eq:stateevolutionSHM}
    q(t) = -\frac{F_0}{2 m \omega^2} \sin \phi \sin(\omega t) + \frac{F_0}{2 m \omega} t \sin (\omega t + \phi) \, .
\end{equation}
The evolution of Eq.\eqref{eq:stateevolutionSHM} has $\phi$ dependence in that 
\begin{equation}
\label{eq:dqbydphi}
\frac{\partial q}{\partial \phi} = -\frac{F_0}{2 m \omega^2} \cos \phi \sin(\omega t) + \frac{F_0}{2 m \omega} t \cos (\omega t + \phi) \, .     
\end{equation}
From Eq.\eqref{eq:dqbydphi} there is, by inspection, no $T$ such that $
\frac{\partial q }{\partial \phi} \big |_{t=T}=0$ for all $\phi$. 
Thus the $q$ cannot be the same for all $\phi$. Similarly, for phase space distributions $\rho_t$ peaked over $q(t)$ the final distribution $\rho_T$ depends on $\phi$ and the DEH condition of Eq.\eqref{eq:DEH} is indeed not respected. The same conclusion holds for initial conditions $q(0)\neq 0$ or $\dot{q} \neq 0$ and off-resonance oscillators (see Supplementary Material I). Thus for this standard model of a harvester, the randomness of the initial phase $\phi$ is passed to the load.

\begin{figure}
    \centering \includegraphics[width=\linewidth]{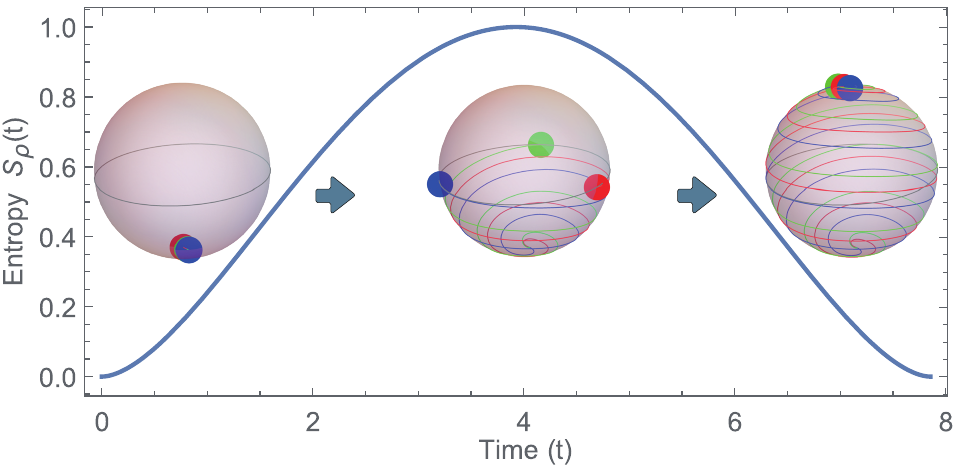}
    \caption{\textbf{Entropy of the state oscillates as the system evolves}. We plot the entropy of the two-level system interacting with the sinusoidal source with uniform random $\phi$ for $A/\omega =0.2$. The spheres show the evolution trajectory for $\phi=0, \frac{2 \pi}{3}, \frac{4 \pi}{3}$ respectively. The initial state $\ket{0}$ and the final state $\ket{1}$ both have entropy 0, while in the middle, the average state, given by Eq.\eqref{eq:rhointermediate}, is a mixed state whose entropy is nonzero.}
    \label{fig:entropy_oscillation}
\end{figure}

\medskip
\textbf{A quantum DEH protocol}. 
There is a well-established method for modelling the coupling between pairs of energy levels of quantum systems under the influence of a sinusoidal potential. 
Both for the case of spins interacting with linearly polarised magnetic fields and 
atoms, whether man-made or artificial, influenced by linearly polarised electric fields, the Hamiltonian $H(t)$ governing the unitary evolution $U(t, t+dt)=\exp(-iH(t)dt)$ is given, in terms of Pauli matrices $X$,$Y$,$Z$ by 
\begin{equation}
\label{eq:Hlinearpolarized}
    H(t) =  -\frac{E}{2} Z + 2 A \cos(\omega t + \phi )X  ,
\end{equation}
where the two-level system has `bare' Hamiltonian $H_0 = - \frac{E}{2} Z$ and 
$2 A \cos(\omega t + \phi )X$ models the influence of the oscillating external field. $E$ is the energy gap of the bare Hamiltonian, and $A$ is a constant that depends on the magnitude of the dipole moment and the oscillation magnitude of the electric field. It is commonly convenient and valid to approximate this Hamiltonian as resulting from a rotating (circularly polarised) field instead, such that (we set $\hbar= 1$) $H(t) \approx H_{RWA}(t)$, where 
\begin{equation}
\label{eq:HRWA}
   H_{RWA}(t) = -\frac{E}{2} Z + A\cos (\omega t + \phi)X + A\sin (\omega t + \phi)Y .
\end{equation}
That rotating wave approximation (RWA) relies on $2A/\omega \ll 1$ (see Supplementary Material II and VII for more details) ~\cite{larson2021jaynes,burgarth2024taming,song2016fast}.
Apart from the RWA approximation, the resonance condition $\omega = E$ will also be demanded.

The Hamiltonian dynamics of Eq.\eqref{eq:HRWA} has an intuitive geometrical interpretation~\cite{ekert2000geometric}. Hermitian operators $O=O^{\dagger}$ like the density matrix $\rho$ and the Hamiltonian $H$ can be represented as real vectors, corresponding to the coefficients in the Pauli basis $\{\id,X,Y,Z\}$: $O
=\frac{1}{2}(\Tr(O\id)\id+\Tr(OX)X+\Tr(OY)Y+\Tr(OZ)Z)$. Often $\Tr(O\id)$ is invariant and only the 3 dimensional so-called Bloch vector $(\Tr(OX) ,\, \Tr(OY),\, \Tr(OZ))^T$ represents $O$. For $O=H_{RWA}$ of Eq.\eqref{eq:HRWA}, the corresponding Bloch vector
\begin{equation}
\label{eq:Rabivec}
\mathbf{\Omega}=(2A \cos (\omega t + \phi), 2A \sin (\omega t + \phi), -E)
\end{equation}
is known as the Rabi vector. The unitary dynamics of the Bloch vector $\mathbf{r}$ corresponding to $\rho$ can be shown from ${\d \rho(t)}/{\d t} = [H, \rho(t)]$ to be
\begin{equation}\label{eq:rotation_equation}
    \frac{\dd \mathbf{r}}{\dd t} = - \mathbf{r} \cross \mathbf{\Omega}\,.
\end{equation}
Thus $\dd \mathbf{\rho}$ is perpendicular to $\mathbf{\rho}$ and $\mathbf{\Omega}$ and can be shown to correspond to a rotation of the state $\mathbf{r}$ around the axis $\mathbf{\Omega}$ with angular velocity  $|\mathbf{\Omega}|$. 

The impact of the dynamics can be seen more easily on a rotating basis. The $Z$-component of $\mathbf{\Omega}$ (Eq.\eqref{eq:Rabivec}) is null if expressing $\mathbf{\Omega}$ in a basis that rotates around the Z-axis (Supplementary Material III):
\begin{equation}
\label{eq:RabivecRotating}
\mathbf{\Omega}_{rot}=(2A \cos \phi, 2A \sin \phi, 0),
\end{equation}
which lies on the equator. Thus, by inspection,  $(0, 0, -1)^T$ representing $\ket{0}$ will evolve to $(0, 0, +1)^T$ representing $\ket{1}$ after time $T= \pi/2A$ irrespective of $\phi$, as illustrated in Fig.~\ref{fig:energy_harvesting_two_phases}. (See Supplementary Material III).

The standard quantum representation yields the same result.
The unitary time evolution operator (detailed derivation in Supplementary Material II)   
\begin{align}
    U_{\phi}(t) =  \cos(A t) \mathbb{I}  - i \sin(A t) X e^{i (\omega t + \phi) Z}\, ,
\end{align}
such that 
\begin{align}
    U_{\phi}(t)\ket{0} =  \cos(A t) \ket{0}  - i \sin(A t)e^{-i (\omega t + \phi)} \ket{1}\, .
\end{align}
Thus when the time $t = \frac{ \pi}{ 2 A}$, the quantum state is guaranteed, up to a global phase, to be the excited state $\ket{1}$, and energy is harvested and stored into the quantum system.  

More details are given in the Supplementary Material. The scheme is shown numerically to be robust to small deviations in the frequency $\omega$, the amplitude $A$, and the stopping time $T$ in Supplementary Material VII. In Supplementary Material IX we give a preliminary derivation estimating the power output for an array of `artificial atom' quantum dots implementing the above DEH scheme.

The entropy of the state, according to the observer who does not know $\phi$,  oscillates. For uniformly random $\phi$,  
\begin{eqnarray}
\label{eq:rhointermediate}
     \rho(t) &=& \int U_{\phi}(t,\phi) \ket{0}\bra{0} U_{\phi}^\dagger(t,\phi) \dd \phi\\
             &=& \cos^2 (At) \ket{0}\bra{0} + \sin^2 (A t) \ket{1} \bra{1} \, .
\end{eqnarray}
Thus the von Neumann entropy of the state, according to someone who does not know $\phi$, is $S_\rho(t) = - \cos^2 (A t) \log \cos^2 (At) -  \sin^2 (A t) \log \sin^2 (At)$ which oscillates from 0 to 1 periodically, as depicted in Fig.~\ref{fig:entropy_oscillation}.

\medskip
{\bf Generalisations.} {The two-level case of DEH under a sinusoidal potential field with random phase can be generalized in several manners. Firstly, higher dimensional systems have 2-level subspaces, such as pairs of electron orbital energy eigenstates, or higher dimensional spin systems. These pairs of energy levels can indeed be selectively pair-wise coupled~\cite{Scheuer_2014,friedrich2006theoretical} such that the above protocol can be directly applied.} 
Secondly, the geometric argument implies that we can relax the restrictions of the source from constant amplitude $2A$ to {\em time-dependent amplitude} $2A(t)$. Eq.\eqref{eq:rotation_equation} and Eq.\eqref{eq:RabivecRotating} imply that the equation of motion is a rotation along an axis on the equator ($\phi$ specifies the direction of the axis) with instantaneous angular speed $2 |A(t)|$. When the angle rotated is $\pi$, the state flips from the south pole to the north pole, irrespective of $\phi$ (see Fig.~\ref{fig:energy_harvesting_two_phases}). Therefore, as long as
\begin{equation}\label{eq:rotation_to_pi}
    \int_0^T A(t) \d t = \pi \, , 
\end{equation}
the protocol works as a DEH. For example, we can ramp up and down the amplitude of the oscillating field, enabling smooth switching on and off the signal (see Supplementary Material VII). 

Thirdly, the sufficiency of Eq.\eqref{eq:rotation_to_pi} can also be used to relax the condition of a single frequency $\omega$. From trigonometry,  if $\frac{\omega_1 + \omega_2}{2}\approx \omega$, then  $A\cos(\omega_1t+\phi)+A\cos(\omega_2t+\phi)\approx 2A \cos\left(\frac{\omega_1 - \omega_2}{2} t\right) \cos\left(\omega t + \phi\right)$ such that we effectively have the same type of source with a time-varying amplitude
$2A \cos\left(\frac{\omega_1 - \omega_2}{2} t\right)$.
Then, from the sufficiency of Eq.\eqref{eq:rotation_to_pi}, a DEH can be constructed by setting $T$ such that $\int_0^T \abs{A \cos (\frac{\omega_1 - \omega _2}{2} t)}\d t = \pi$. Finally, for every bare Hamiltonian $H_0$ one can find a family of external Hamiltonians such that DEH is achieved (see Supplementary Material VIII).

\medskip
\textbf{Quantum-inspired classical DEH protocol.} 
The geometrical interpretation of the quantum protocol inspires us to consider whether classical rotating objects may achieve DEH. 

The angular momentum of a charged sphere in an electric field can essentially obey the same evolution equation as Eq.\eqref{eq:rotation_equation}. We now give an outline of the argument with details provided in Supplementary Material V for the electric dipole (and VI for the magnetic dipole). Suppose the charge distributed on the sphere is non-uniform, forming a dipole with dipole moment $\mathbf{d}$ and that the mass of the sphere is uniform with a moment of inertia $I$ for any rotation axis through the sphere's center. An external electric field $\mathbf{E}$ exerts a torque on the dipole: $\tau = \mathbf{d} \times \mathbf{E}$. 
The sphere rotates along the axis $\hat{\mathbf{d}} = {\mathbf{d}}/{|\mathbf{d}|}$ with angular momentum $\mathbf{L} = (L_x, L_y, L_z) = \alpha {\mathbf{d}}$, where $\alpha$ is the proportional constant. Thus
\begin{equation}
\label{eq:classicalLevolution}
    \frac{\d \mathbf{L}}{\d t} = - \frac{1}{\alpha} \mathbf{E} \times \mathbf{L} \, ,
\end{equation}
which is essentially the same as Eq.\eqref{eq:rotation_equation}. 

This classical charged sphere system can achieve DEH for an oscillating potential. The Hamiltonian of the rotating charged sphere $H = {|\mathbf{L}|^2}/{2 I} - \mathbf{d} \cdot \mathbf{E}$ with $-\mathbf{d} \cdot \mathbf{E}$ the potential. Suppose $\mathbf{E} = {E}_0 \hat{\mathbf{z}} + 2 A \cos (\omega t + \phi ) \hat{\mathbf{x}}$, where $E_0$ is a constant electric field that with angular frequency $\omega$. Then the evolution of the rotating sphere follows the same equation as the quantum state given Eq.(\ref{eq:rotation_equation}) and Eq.(\ref{eq:RabivecRotating}), and thus flips the dipole deterministically after $T= \pi/2A$. 

Other classical examples are expected to exist. A variant of Eq.\eqref{eq:classicalLevolution} can also emerge when describing the magnetization vector in magnetic resonance (see Supplementary Material IV). Moreover, the universality of Hamiltonian mechanics~\cite{goldstein2011classical}, a universality expressed in mechatronics via the bond graph formalism~\cite{karnopp2012system}, implies that other physical scenarios can respect the same mathematics. 

The results of this section suggest that quantum-inspired approaches using classical technology can achieve an advantage in energy harvesting output over other classical techniques. Classical technology is more technologically mature, though a reason for using quantum systems may be the suppression of chaos as exemplified by the suppression of diffusion following kicks to rotating objects~\cite{FishmanSG82}.

\medskip
\textbf{DEH consistent with 2nd law and Liouville's theorem}. 
DEH does not violate the 2nd law of thermodynamics. A DEH protocol can cyclically extract work from a {\em partly random source} without generating any other effect. The stochastic profile of the source here is not white Gaussian as in the standard model of thermal electromagnetic radiation~\cite{johnson1928thermal, nyquist1928thermal, swati2022quantifying}; in particular the autocorrelation $\langle E(t)E(t+\tau)\rangle=0$ for thermal radiation~\cite{ou2017quantum} which is not the case here. The second law of thermodynamics~\cite{boltzmann1974second, lieb1999physics} states that work cannot be extracted from a single {\em heat bath} without producing any other effect.

DEH does not violate the reversibility of classical and quantum dynamics. In classical dynamics one version of this statement is Liouville's theorem~\cite{goldstein2011classical}, that under an infinitessimal update of $t$,$x$ and $p$ the probability density is invariant, $\d \rho(x,p,t)=(\partial_x\rho) \d x+(\partial_p\rho) \d p+(\partial_t\rho) \d t=0$, which captures the intuition that the time evolution permutes the phase space points around without merging any trajectories. In phase space quantum theory~\cite{zachos2005quantum}, that specific equation may be violated in a certain sense by non-local `jumping' in phase space~\cite{jiang2024framework} but any unitary evolution $U$ is nevertheless manifestly reversible by $U^{\dagger}$ and thus not many-to-one: $U^{\dagger}U\ket{\psi}=\ket{\psi}$. Reversible dynamics, whether quantum or classical, is moreover entropy preserving. The DEH protocol is outside of the domain of validity of the above statements, because it involves the dynamics of the system under {\em different} Hamiltonians. A probabilistic combination of different Hamiltonians may, as seen above, implement different trajectories that all lead to a fixed state at a given time and to oscillating entropy.

\medskip
\textbf{Summary and Outlook.} We formulate and present a novel property of a quantum energy harvester---the ability to harvest energy from a fluctuating source without passing its randomness to the load.  A two-level quantum system can deterministically harvest energy from a sinusoidal fluctuating source without being affected by its random initial phase. We also designed a classical DEH protocol, exploiting angular momentum. 

Several possible developments would be valuable: (i) generalize the DEH quantum protocol to more types of random sources and quantum systems, (ii) application to  efficient wireless power transmission, (iii) application to AC-to-DC rectification e.g.\ for overcoming the rectification problem of rectenna~\cite{surender2021rectenna}, and to quantum rectifiers~\cite{vidan2004triple}
(iv) implementation in spintronics~\cite{vzutic2004spintronics}
(v) application to charging quantum batteries~\cite{campaioli2017enhancing,ferraro2018high,song2024remote}, and spin batteries~\cite{brataas2002spin, wang2004spin}, (vi) formulate a theory for energy harvesting that applies both to quantum and classical mechanics, along the lines of recent results on generalised probabilistic theory mechanics\cite{JiangTD2024a, JiangTD2024b, PlavalaK2022, LinD2020}, (vii) inclusion of DEH into the resource theory of transient voltage sources~\cite{swati2022quantifying}. 

\textbf{Acknowledgments.} We gratefully acknowledge discussions with Huaiyang Yuan, Xun Gao, Guanjie He, Xin Wang, and Sen Yang. We acknowledge support from the National Natural Science Foundation of China (Grants No. 12050410246, No.1200509, No.12050410245) and the City University of Hong Kong (Project No. 9610623).

\clearpage

\onecolumngrid
\begin{center}
	\textbf{\large Supplementary Materials for Quantum harvester enables energy transfer without randomness transfer or
dissipation}\\
	 \vspace{2ex}
\end{center}

\section{Classical linear oscillator fails to implement DEH.}
\label{SM:I}
In this section, we present the full analysis of why a classical linear oscillator fails to be a DEH protocol for the sinusoidal source with a random initial phase.

We divide the analysis into two subsections; one deals with resonant oscillators, and the other deals with off-resonant oscillators.

\medskip
\textbf{At resonance.}
Suppose that the bare Hamiltonian of the harmonic oscillator is 
\begin{equation}
H_0 = \frac{p^2}{2m} + \frac{kq^2}{2},
\end{equation}
where \( m \) is the mass and \( k \) is the spring constant. The system experiences an oscillating external potential 
\begin{equation}
V(t) = -F_0 \cos (\omega t + \phi) q,
\end{equation}
where $F_0$ is the maximal force.

The equation of motion for this system can be derived from Newton's second law, given by
\begin{equation}\label{eq:equation_of_motion_linear_oscillator}
m \ddot{q} + k q = F_0 \cos (\omega t + \phi).
\end{equation}
This is a non-homogeneous second-order linear differential equation.
First, we find the homogeneous solution of the corresponding homogeneous equation:
   \begin{equation}
   m \ddot{q} + k q = 0,
   \end{equation}
whose general solution is:
   \begin{equation}
   q_h(t) = A \cos(\omega_0 t) + B \sin(\omega_0 t),
   \end{equation}
   where \( \omega_0 = \sqrt{\frac{k}{m}} \) is the natural frequency of the oscillator. At resonance, we have $\omega_0 = \omega$.

Now we need to find the particular solution for the inhomogeneous equation. We guess:
   \begin{equation}
   q_p(t) =  \left(C \sin(\omega t + \phi) \right) t.
   \end{equation}
   Substituting it into the differential equation Eq.\eqref{eq:equation_of_motion_linear_oscillator} and we get $C = \frac{F_0}{2 m \omega }$.
Thus, the general solution is:
\begin{equation}
q(t) = A \cos(\omega t) + B \sin(\omega t) + \frac{F_0}{2 m \omega} t \sin(\omega t + \phi)
\end{equation}
Given \( q(0) = 0 \) and \( \dot{q}(0) = 0 \), we solve for \( A \) and \( B \) and get \( A = 0 \) and \( B = -\frac{F_0}{2 m \omega^2} \sin \phi \). Therefore, the equation of motion for the linear oscillator at resonance is
\begin{equation}
q(t) = -\frac{F_0}{2 m \omega^2} \sin \phi \sin(\omega t) + \frac{F_0}{2 m \omega} t \sin (\omega t + \phi) \, ,
\end{equation}
which justifies the solution we present in the main text. Since
\begin{equation}
\frac{\partial q}{\partial \phi} = -\frac{F_0}{2 m \omega^2} \cos \phi \sin(\omega t) + \frac{F_0}{2 m \omega} t \cos (\omega t + \phi) \, ,     
\end{equation}
there is no $T$ such that $
\frac{\partial q }{\partial \phi} \big |_{t=T}=0$ for all $\phi$. Thus, the final state cannot be the same for all $\phi$, and randomness of the initial phase $\phi$ is passed to the load.

If, however, we start with a more general initial configuration, $q(0)=\alpha$ and $\dot{q}(0) = \beta$, then it still cannot achieve DEH. For this general initial configuration, the equation of motion for the linear oscillator at resonance is 
\begin{equation}
    q(t) = \alpha \cos (\omega t) + [\frac{\beta}{\omega} - \frac{F_0}{2 m \omega^2} \sin \phi ]\sin (\omega t) + \frac{F_0}{2 m \omega} t \sin (\omega t + \phi) \, .
\end{equation}
Then again, we have
\begin{equation}
    \frac{\partial q}{\partial \phi} = \frac{\partial q}{\partial \phi} = -\frac{F_0}{2 m \omega^2} \cos \phi \sin(\omega t) + \frac{F_0}{2 m \omega} t \cos (\omega t + \phi) \, ,     
\end{equation}
and for every $\alpha$ and $\beta$ there is no $T$ such that $
\frac{\partial q }{\partial \phi} \big |_{t=T}=0$ for all $\phi$. It means no matter what the initial configuration is, the linear oscillator at resonance cannot implement a DEH.

\medskip
\textbf{Off-resonance.}
Now, let us solve the equation of motion for the off-resonance regime and prove that it cannot achieve DEH either. To find the solution, we need to solve for one particular solution for Eq.~\eqref{eq:equation_of_motion_linear_oscillator}. We guess the particular solution is
\begin{equation}
    q_p(t) = C \cos (\omega t + \phi ) \, .
\end{equation}
Substituting $q_p(t)$ into the differential equation Eq.\eqref{eq:equation_of_motion_linear_oscillator} and we have $C = \frac{F_0}{m (\omega_0^2 - \omega^2})$. Thus the general solution for Eq.~\eqref{eq:equation_of_motion_linear_oscillator} is
\begin{equation}
    q(t) = A \cos (\omega_0 t ) + B \sin (\omega_0 t ) + \frac{F_0}{m (\omega_0^2 - \omega^2)} \cos (\omega t + \phi) \, .
\end{equation}
Now suppose the initial condition is $q(0) = \alpha $ and $\dot{q}(0) = \beta$, then we can solve for $A = \alpha -\frac{F_0 \cos \phi}{m (\omega_0^2 - \omega^2) }$ and $B = \frac{\beta}{\omega_0} + \frac{F_0 \omega \sin \phi}{\omega_0 m(\omega_0^2 - \omega^2)}$. Substituting $A$ and $B$ into the solution we have
\begin{equation}
    q(t) = \alpha \cos (\omega_0 t ) + \frac{\beta}{\omega_0} \sin (\omega_0 t) + \frac{F_0 \cos \phi}{m (\omega_0^2 - \omega^2)}[\cos (\omega t ) - \cos (\omega_0 t)] + \frac{F_0 \sin \phi}{m (\omega_0^2 - \omega^2)} [\frac{\omega}{\omega_0} \sin \omega_0 t - \sin (\omega t )]
\end{equation}
and 
\begin{equation}
    \dot{q}(t) =- \alpha \omega_0 \sin (\omega_0 t ) + {\beta} \cos (\omega_0 t) + \frac{F_0 \omega_0 \cos \phi}{m  (\omega_0^2 - \omega^2)}[\frac{\omega}{\omega_0} \sin (\omega t ) -  \sin (\omega_0 t)] + \frac{F_0 \omega \sin \phi}{m (\omega_0^2 - \omega^2)} [\cos \omega_0 t - \cos (\omega t )] .
\end{equation}
Take the partial derivative with respect to $\phi$, and we have
\begin{equation}
    \frac{\partial q(t)}{\partial \phi} =- \frac{F_0 \sin \phi}{m (\omega_0^2 - \omega^2)}[\cos (\omega t ) - \cos (\omega_0 t)] + \frac{F_0 \cos \phi}{m (\omega_0^2 - \omega^2)} [\frac{\omega}{\omega_0} \sin (\omega_0 t) - \sin (\omega t )] ,
\end{equation}
and
\begin{equation}
    \frac{\partial \dot{q}(t)}{\partial \phi}  =-\frac{F_0 \omega_0 \sin \phi}{m  (\omega_0^2 - \omega^2)}[\frac{\omega}{\omega_0} \sin (\omega t ) -  \sin (\omega_0 t)] + \frac{F_0 \omega \cos \phi}{m (\omega_0^2 - \omega^2)} [\cos (\omega_0 t) - \cos (\omega t )].
\end{equation}
We need to guarantee $\frac{\partial q(t)}{\partial \phi} = 0$ and $\frac{\partial \dot {q}(t)}{\partial \phi} = 0 $ for all $\phi$ when $t=T$, which is equivalent to 
\begin{equation}
    \cos (\omega T) - \cos (\omega_0 T) =0 \,\,\, \text{and}\,\,\, \frac{\omega}{\omega_0} \sin (\omega t) - \sin (\omega_0 t) =0  \,\,\, \text{and}\,\,\,  \sin (\omega t) - \frac{\omega}{\omega_0} \sin (\omega_0 t)    = 0 \, .
\end{equation}
In order to simultaneously satisfy $\frac{\omega}{\omega_0} \sin (\omega T) - \sin (\omega_0 T) =0$ and $\sin (\omega T) - \frac{\omega}{\omega_0} \sin (\omega_0 T)    = 0$, we must have $\sin (\omega T) = \sin (\omega_0 T) = 0$. Therefore, in order to satisfy these constraints, we must have
\begin{equation}
    \cos (\omega T) = \cos (\omega_0 T ) = \pm 1 \,\,\, \text{and}\,\,\, \sin (\omega T) = \sin (\omega_0 T) = 0 \, .
\end{equation}
These equations can be satisfied, for example, when $\omega_0 =2$, $\omega=1$ and $T= 2\pi$. However, as long as the above constraints are satisfied, we automatically have
\begin{equation}
    q(T) = \pm q(0) \, \, \, \text{and} \,\,\, \dot{q}(T) = \pm \dot{q} (0) \, .
\end{equation}
This implies
\begin{equation}
    \Delta E = [\frac{1}{2}k q(T)^2 + \frac{1}{2} m \dot{q}(T)^2 ]- [\frac{1}{2}k q(0)^2 + \frac{1}{2} m \dot{q}(0)^2] = 0 \, ,
\end{equation}
which contradicts the condition of non-zero energy harvesting. Either there is energy transferred, but the randomness of the initial phase is also transferred, or  
there is no randomness transfer but there is also no energy harvested.

Thus, we conclude that the linear oscillator cannot implement DEH, no matter the initial condition and no matter if it is at resonance or not.

\section{Solving Rabi oscillation of a quantum dipole moment.}
\label{SM:II}
 A quantum dipole modelled as a two-level system with bare Hamiltonian $H_0 = - \frac{E}{2} Z$ has a total Hamiltonian
\begin{equation}
    H(t) =  -\frac{E}{2} Z + 2 A \cos(\omega t + \phi )X  ,
\end{equation}
when interacting with the fluctuating electric field, where $E$ is the energy gap, $A$ is a constant that depends on the magnitude of the dipole moment and the oscillation magnitude of the electric field.

At the resonant condition $\hbar \omega = E $ (assuming $\hbar=1$ for simplicity), apply the rotating wave approximation ($\frac{2A}{\omega} \ll 1$, then $2\cos (\omega t+ \phi) X
\approx 
\begin{bmatrix}
0  &  e^{i(\omega t+\phi) }\\
 e^{-i(\omega t+\phi) } & 0
\end{bmatrix}$) and we have the following total Hamiltonian,
\begin{equation}
    \Tilde{H} = -\frac{E}{2} Z +  \begin{pmatrix}
0 &  A e^{i(\omega t + \phi)} \\
 A e^{-i(\omega t + \phi)} & 0
\end{pmatrix} .
\end{equation}
In the interaction picture, set 
\begin{equation}
    U_0 (t) =\begin{pmatrix}
e^{i \frac{E t + \phi}{2}} & 0 \\
0 & e^{-i \frac{E t + \phi}{2}}
\end{pmatrix}\, .
\end{equation}
Then the effective Hamiltonian in the interaction picture is
\begin{equation}
    V = U_0^\dagger (t) \begin{pmatrix}
0 &  A e^{i(\omega t + \phi)} \\
 A e^{-i(\omega t + \phi)} & 0
\end{pmatrix}  U_0(t)=  A X\, .
\end{equation}
Then the time evolution operator in the interaction picture is 
\begin{equation}
    U_{I}(t) = e^{- i A t X}  = \cos(A t) \mathbb{I} - i \sin(A t) X \, .
\end{equation}
Combined together, we have the time evolution operator in the Schr\"{o}dinger picture as
\begin{align}
    U(t, \phi) & = U_0^\dagger(t) U_{I} U_0(t) =  \cos(A t) \mathbb{I}  - i \sin(A t) X e^{i (\omega t + \phi) Z}\, .
\end{align}

\section{Geometric interpretation of the dynamics of a two-level system: Bloch vector and Rabi vector.}
\label{SM:III}
In this section, we show that the Rabi oscillation that is induced by a rotating field on a two-level system has an interesting property---the population transfer from the ground state to the excited state is independent of the phase of the rotating field. This property is the key mechanism for energy transfer without randomness transfer using a quantum harvester. We retrieve derivations used in Ref.~\cite{ekert2000geometric}.

A two-level quantum system can be described by the following density operator,
\begin{equation}
    \rho = \frac{1}{2}(\mathbb{I} + \mathbf{r} \cdot \sigma) = \frac{1}{2} \begin{pmatrix}
        1 + r_z & r_x - i r_y\\
        r_x + i r_y & 1- r_z
    \end{pmatrix} \, ,
\end{equation}
with a real vector $\mathbf{r} = (r_x, r_y, r_z)$ which is called the Bloch vector and $\sigma$ is the vector formed by the Pauli matrices. And any $2\times2$ Hamiltonian can be expressed as
\begin{equation}
    H = \frac{\hbar}{2} (\Omega_0 \mathbb{I} + \mathbf{\Omega} \cdot \sigma ) \, ,
\end{equation}
with $\mathbf{\Omega}$ being the Rabi vector.

The equation of motion for state $\rho$ under the evolution of this Hamiltonian $H$ is given by
\begin{equation}
    i \hbar \frac{\dd }{\dd t} \rho = [H, \rho] \, .
\end{equation}
Because of the following identify
\begin{equation}
    (\mathbf{a} \cdot \sigma) (\mathbf{b} \cdot \sigma) = (\mathbf{a}  \cdot \mathbf{b}) \mathbb{I} + i (\mathbf{a} \times \mathbf{ b} \cdot \sigma ) \, ,
\end{equation}
we have the equation of motion expressed in terms of the Bloch vector and the Rabi vector,
\begin{equation}
    \frac{\dd }{\dd t} \mathbf{r} = \mathbf{\Omega \times r} \, . \label{eq:rotation}
\end{equation}
For a spin in a rotating magnetic field or a dipole moment in a rotating electric field, the Hamiltonian is
\begin{equation}
    H(t) = \frac{\hbar}{2} \begin{pmatrix}
        \omega_0 & \omega_1 e^{- i (\omega t +\phi)} \\
        \omega_1 e^{i (\omega t + \phi)} & - \omega_0 
    \end{pmatrix} \, , \label{eq:hamiltonian}
\end{equation}
with $\phi$ the initial phase of the oscillating field, $\hbar \omega_0$ system's intrinsic energy gape (which we call $E$ in the previous section and in the main text), $\omega$ is the angular frequency of the rotating filed, $\hbar \omega_1$ (which we call $2A$ in the previous section and the main text) is the amplitude of the rotating field, respectively.
For a spin in an oscillating magnetic field (see Sec.~\ref{SM:VI}), or for a dipole moment in an oscillating electric field (see Sec.~\ref{SM:V}), the Hamiltonian after rotating wave approximation (RWA) is also Eq.\eqref{eq:hamiltonian}.
This Hamiltonian gives the following Rabi vector,
\begin{equation}
    \Omega_x = \omega_1 \cos (\omega t + \phi) ,\, \,\, \Omega_y = \omega_1 \sin(\omega t + \phi), \,\,\, \Omega_z = \omega_0 \, .
\end{equation}
In the interaction picture, we change the equation in a rotating reference frame, which makes it more convenient to solve Eq.~\eqref{eq:rotation}. Specifically, we have
\begin{equation}
    \mathbf{r}(t) = R_z (\omega t ) \mathbf{r}'(t) \, ,
    \, \, \, 
    \mathbf{\Omega}(t) = R_z(\omega t) \mathbf{\Omega}'(t) \, 
\end{equation}
where $R_z(\omega t)$ is defined as 
\begin{equation}
 R_z(\omega t) = \begin{pmatrix}
     \cos (\omega t) & - \sin (\omega t) & 0 \\
     \sin(\omega t) & \cos (\omega t) & 0 \\
     0 & 0& 1 
 \end{pmatrix} \, .   
\end{equation}
Therefore, we have 
\begin{equation}
    \frac{\dd }{\dd t} \mathbf{r}' = \mathbf{\Omega}' \times \mathbf{r}' \, ,
\end{equation}
where the new Rabi vector is time-independent 
$\mathbf{\Omega}'$,
\begin{equation}
 \Omega_x' = \omega_1 \cos (\phi), \, \, \, \Omega_y'= \omega_1 \sin (\phi), \, \, \, \Omega_z' = \omega_0 - \omega \, .   
\end{equation}
Then, at resonance, $\omega_0 = \omega$, the new Rabi vector $\mathbf{\Omega}'$ is on the equator of the Bloch sphere. By symmetry, no matter the value of $\phi$, the rotation can move the ground state $\ket{g}$ to $\ket{e}$ after $T=\pi/\omega_1$.

\section{Classical Spin Dynamics implements DEH.} \label{SM:IV}

In this section, we show that the magnetization of a material at magnetic resonance can, in principle, implement DEH. We will show that the magnetization follows the same equation of motion driven by an external magnetic field, assuming there is negligible dissipation.

The dynamics of the magnetization vector $\mathbf{r}$ under magnetic fields and damping is described by the Landau-Lifshitz-Gilbert (LLG) equation~\cite{lakshmanan2011fascinating}, which is expressed as,
\begin{equation}
    \frac{d\mathbf{r}}{dt} = -\gamma \mathbf{r} \times \mathbf{B}_{\text{eff}} + \frac{\alpha}{r_s} \mathbf{r} \times \frac{d\mathbf{r}}{dt}
\end{equation}
where $\mathbf{r}$ is the magnetization vector, $\gamma$ is the gyromagnetic ratio, $\mathbf{B}_{\text{eff}}$ is the effective magnetic field, including external and internal contributions,  $\alpha$ is the dimensionless damping coefficient and $r_s$ is the saturation magnetization of the material.
The equation combines two main terms:  The precessional term $(-\gamma \mathbf{r} \times \mathbf{B}_{\text{eff}})$, which describes the precession of $\mathbf{r}$ around the effective field. And the damping term $(\frac{\alpha}{r_s} \mathbf{r} \times \frac{d\mathbf{r}}{dt})$, introduced by Gilbert, which models the energy dissipation and stabilizes the magnetization.

Now for material with negligible damping coefficient, the equation of motion for the magnetization is
\begin{equation}
    \frac{d\mathbf{r}}{dt} = -\gamma \mathbf{r} \times \mathbf{B}_{\text{eff}} \, .
\end{equation}

Magnetization is arguably a macroscopic quantum effect since magnetization originates from the electron's spin and orbital motion. Nevertheless, we can construct a classical protocol via a classical rotating sphere, as shown in the following sections.

\section{DEH via a charged rotating sphere (electric dipole).}
\label{SM:V}

In this section, we describe a classical DEH protocol for harvesting energy from an electric field. The protocol uses a charged rotating sphere with a fixed electric dipole as the harvester. 

Consider a sphere with uniform mass and it has a moment of inertia $I$ for all the rotation axes. Then assume there are charges distributed on the sphere such that it has a dipole moment $\mathbf{d}$. This can be achieved, for example, by putting a positive charge at the north pole and put an equivalent amount of negative charge at the south pole. Now suppose the sphere rotates along the axis $\hat{\mathbf{d}} = {\mathbf{d}}/{|\mathbf{d}|}$ with angular momentum $\mathbf{L} = (L_x, L_y, L_z) = \alpha {\mathbf{d}}$, where $\alpha$ is the proportional constant. Now put the rotating charged sphere in an electric field, and it has a Hamiltonian
\begin{equation}
    H = \frac{|\mathbf{L}|^2}{2 I} - \mathbf{d} \cdot \mathbf{E} = \frac{|\mathbf{L}|^2}{2 I} - \frac{1}{\alpha} \mathbf{L} \cdot \mathbf{E} \, .
\end{equation}

Then by the Hamiltonian equation, each components of the angular momentum \(\mathbf{L}\) are given by
\begin{equation}
    \frac{dL_i}{dt} = \{ L_i, H \},
\end{equation}
where \( \{ \cdot, \cdot \} \) denotes the Poisson bracket. For angular momentum components, the Poisson brackets satisfy the relation,
\begin{equation}
    \{ L_i, L_j \} = \epsilon_{ijk} L_k \, .
\end{equation}
By the chain rule for Poisson brackets, we have:
\begin{equation}
\{ L_i, H \} = \sum_{j} \frac{\partial H}{\partial L_j} \{ L_i, L_j \}.
\end{equation}
Here, \(\frac{\partial H}{\partial L_j}\) is the partial derivative of \(H\) with respect to \(L_j\).
Then we compute the time evolution of \(\mathbf{L}\):
\begin{equation}
    \frac{dL_i}{dt} = \{ L_i, H \} = - \sum_{j,k} \epsilon_{ijk} L_j \frac{\partial H}{\partial L_k}.
\end{equation}
The partial derivative of the Hamiltonian with respect to \(L_k\) is
\begin{equation}
\frac{\partial H}{\partial L_k} = \frac{\partial}{\partial L_k} \left( \frac{\mathbf{L}^2}{2I} - \frac{1}{\alpha} \mathbf{L} \cdot \mathbf{E}(t) \right) = \frac{L_k}{I} - \alpha E_k(t).
\end{equation}

Thus, the equation of motion becomes
\begin{equation}
\frac{dL_i}{dt} = - \sum_{j,k} \epsilon_{ijk} L_j \left( \frac{L_k}{I} - \frac{1}{\alpha} E_k(t) \right).
\end{equation}
This equation can be written as
\begin{equation}
\frac{d\mathbf{L}}{dt} = - \frac{1}{I} \mathbf{L} \times \mathbf{L} + \frac{1}{\alpha} \mathbf{L} \times \mathbf{E}(t),
\end{equation}
since the \(i\)-th component of the cross product \(\mathbf{L} \times \mathbf{E}\) is given by $ (\mathbf{L} \times \mathbf{E})_i = \sum_{j,k} \epsilon_{ijk} L_j E_k$. 
Furthermore, because \(\mathbf{L} \times \mathbf{L} = 0\), we simplify the equation of motion for the angular momentum of the sphere to
\begin{equation}
    \frac{\d \mathbf{L}}{\d t} = - \frac{1}{\alpha} \mathbf{E} \times \mathbf{L} \, .
\end{equation}

\section{DEH via a charged rotating sphere (magnetic dipole).}
\label{SM:VI}

In this section, we describe a classical DEH protocol for harvesting energy from a magnetic field. The protocol uses a charged rotating sphere with a fixed magnetic dipole as the harvester.

Consider a charged rotating sphere with moment of inertia \( I \) and angular momentum \( \mathbf{L} = (L_x, L_y, L_z) \). The magnetic moment \(\boldsymbol{\mu}\) of this rotating charged sphere is proportional to the angular momentum, given by $ 
\boldsymbol{\mu} = \beta \mathbf{L},
$,
where \( \beta \) is a proportionality constant. Since the Hamiltonian of the magnetic dipole moment $\boldsymbol{\mu}$ is $- \boldsymbol{\mu} \cdot \mathbf{B}$, then the Hamiltonian of the system in the presence of a time-dependent external magnetic field \(\mathbf{B}(t)\) is given by
\begin{equation}
   H = \frac{\mathbf{L}^2}{2I} - \beta \mathbf{L} \cdot \mathbf{B}(t). 
\end{equation}
The Hamiltonian equations of motion can be similarly solved as we did for the rotating electric dipole previously,
\begin{equation}
\frac{d\mathbf{L}}{dt} =  \beta \mathbf{L} \times \mathbf{B}(t).
\end{equation}

This is the equation of motion for a charged rotating sphere in a time-dependent magnetic field. It follows the same equation as the magnetic spin evolution, so the same protocol can be used to flip this charged rotating sphere similarly to the rotating electric dipole case in Sec.~\ref{SM:V}.

\section{Numerical results for the robustness of the quantum DEH protocol.}
\label{SM:VII}
In this section, we investigate the robustness of the quantum DEH protocol against parameter shifts.
Specifically, we numerically test the population of the excited states when the amplitude $A$, frequency $\omega$, and stopping time $T$ have small deviations. We also tested the effect of pulse shaping on the accuracy of our protocol.

Our simulation results demonstrate that the quantum DEH protocol is robust against small errors in the external field frequency, amplitude, and protocol's stopping time. This analysis suggests that our protocol has great potential to be applied in realistic scenarios where there exists noise and errors.

\begin{figure}[b]
    \centering
    \includegraphics[width = 0.49 \linewidth]{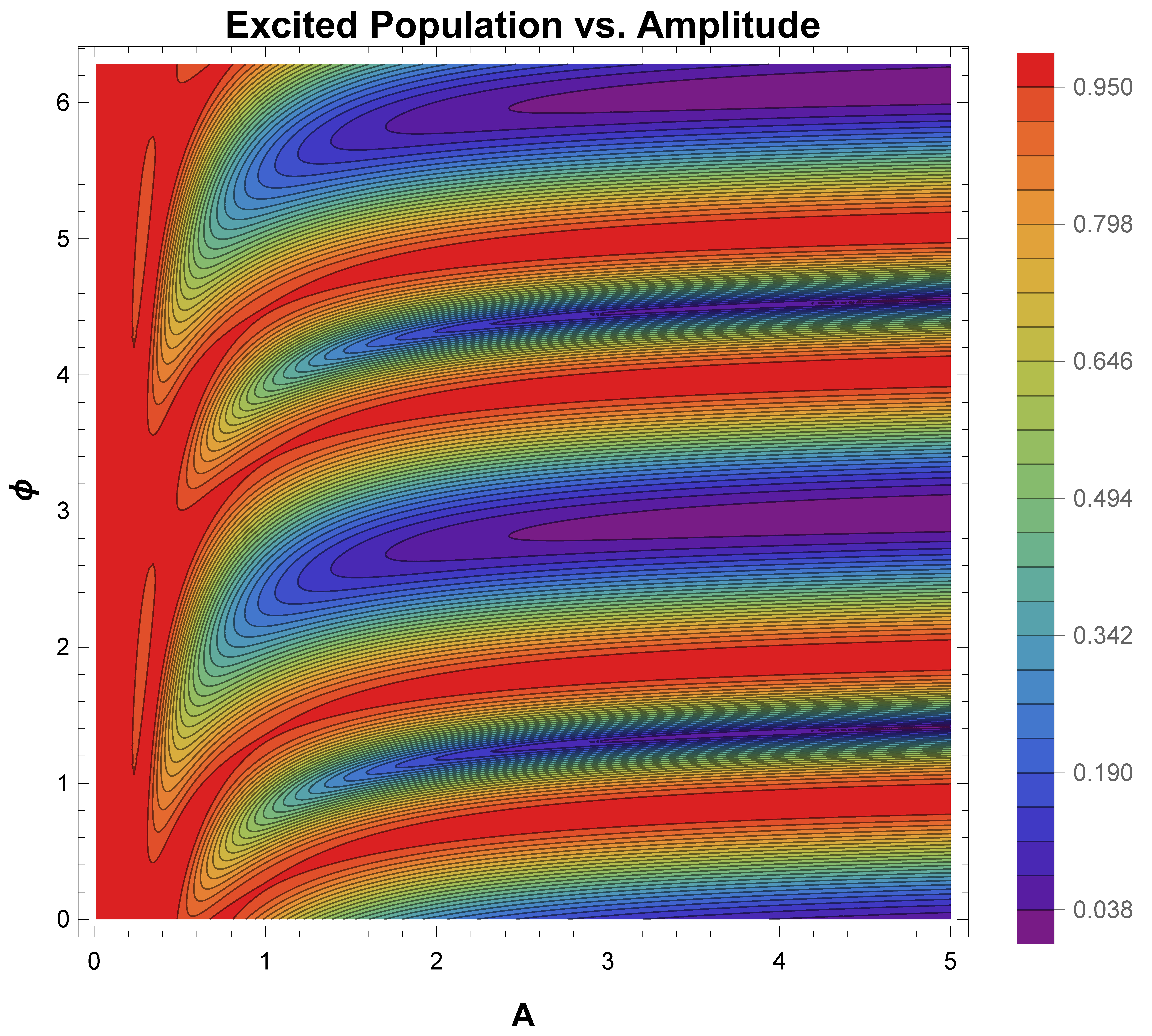}
    \includegraphics[width = 0.49 \linewidth]{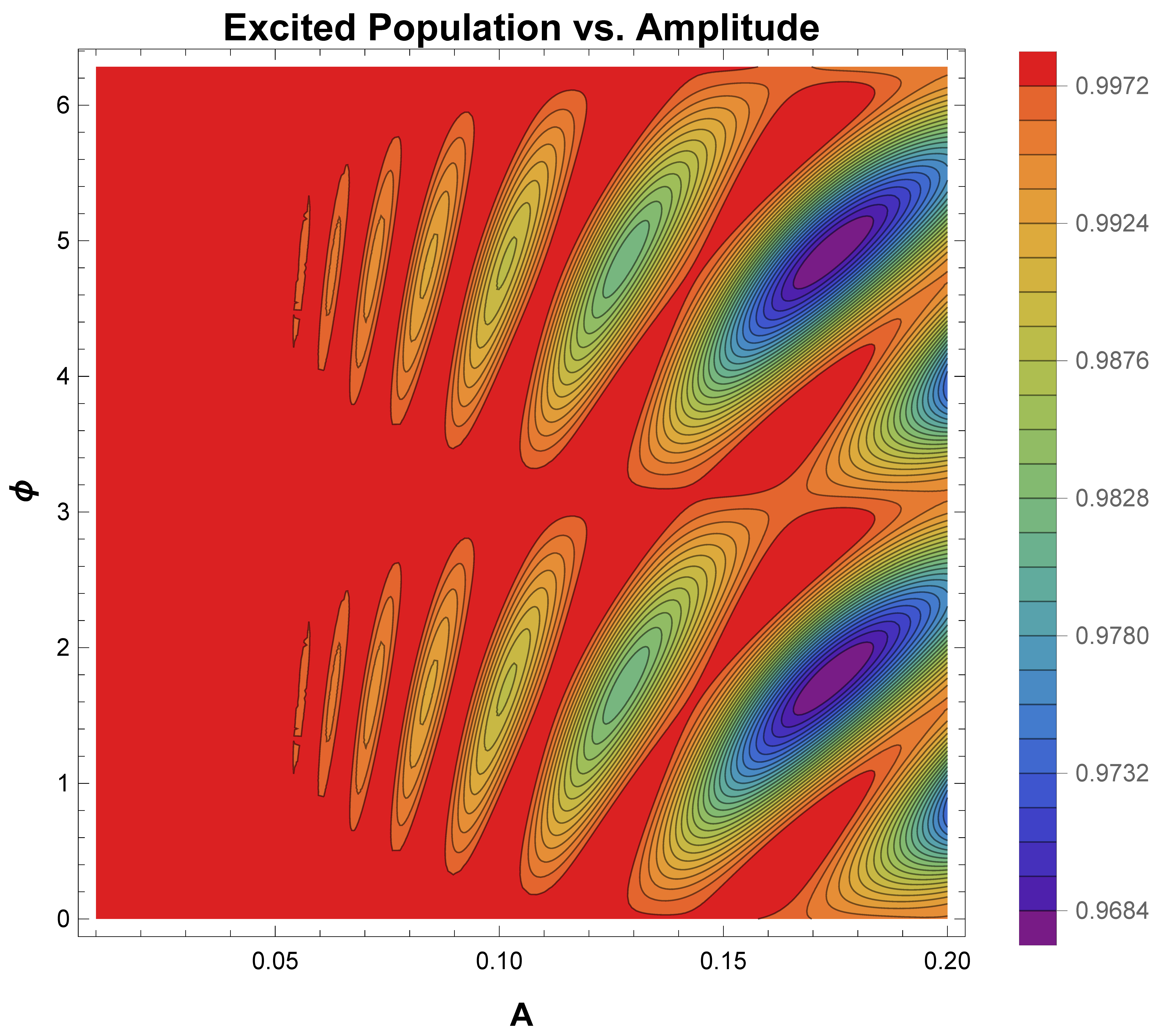}
    \caption{\textbf{Independence of the initial phase $\phi$ for small interaction strength.} Both figures plot the population on the excited state when interacting with an external potential. The total Hamiltonian is $H = -
    \frac{E}{2} Z + 2 A \cos (\omega t + \phi ) X$ (both plots assume $E=1$). The two figures plot the population versus various interaction strengths $A$ and initial phase $\phi$. While the independence on $\phi$ fails to hold for strong interaction (approximately $A>0.2$), it holds quite nicely (up to 0.9972) for weak interaction (approximately $A<0.05$). }
    \label{fig:interaction_strength}
\end{figure}

\medskip
\textbf{Independence on the initial phase $\phi$ for weak interaction.}
The crucial feature of our proposed quantum DEH protocol---the ability to excite the ground state to the excited state regardless of the external field's initial phase---in fact, depends on the validity of the rotating wave approximation. For sufficiently weak interaction, the protocol has negligible dependence on the random initial phase $\phi$ and, therefore, implements a DEH. We numerically analyze the independence on the phase $\phi$ for various interaction strength $A$.
Assuming the total Hamiltonian is $H = -\frac{E}{2} Z + 2 A \cos (\omega t + \phi ) X$ with $E=1$. In Fig.~\ref{fig:interaction_strength}, we plot the population versus various interaction strengths $A$ and initial phase $\phi$.
It can be seen from the figure that the protocol's performance is independent of the phase $\phi$ for sufficiently small interaction strength $A$. Specifically, the independence holds quite well (with a probability of at least 0.9972 to excite the state) for weak interaction with $A<0.05$, and if the tolerance is relaxed to a success probability of excitation at least 0.95, the interaction strength can be extended to $A <0.2$.

\begin{figure}
    \centering
    \includegraphics[width=0.49\linewidth]{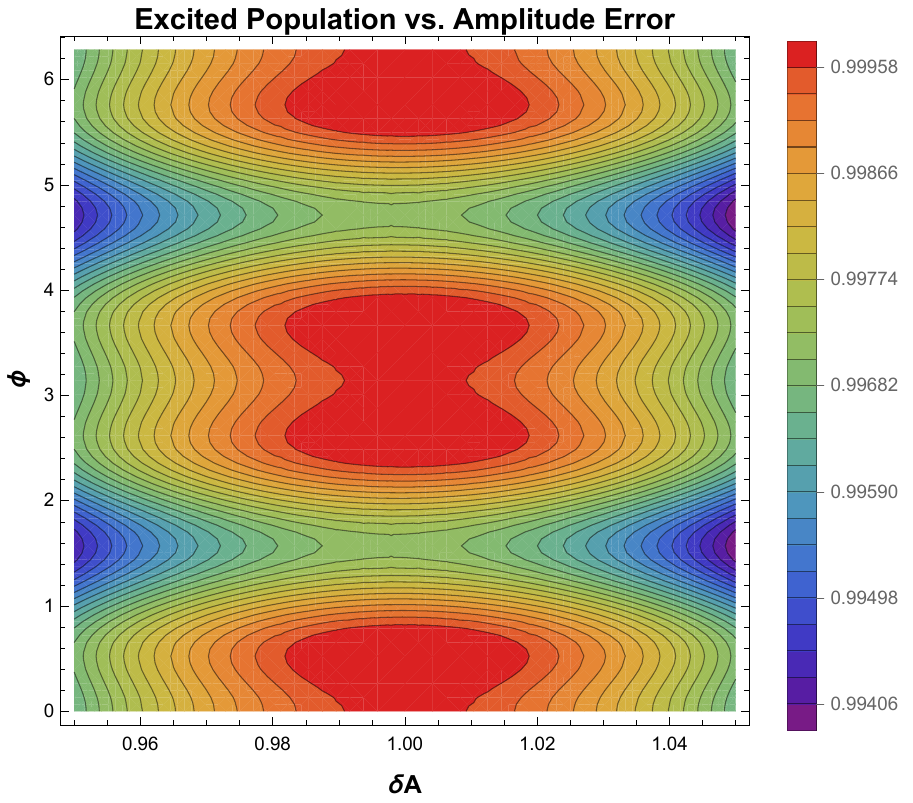}
    \includegraphics[width=0.49\linewidth]{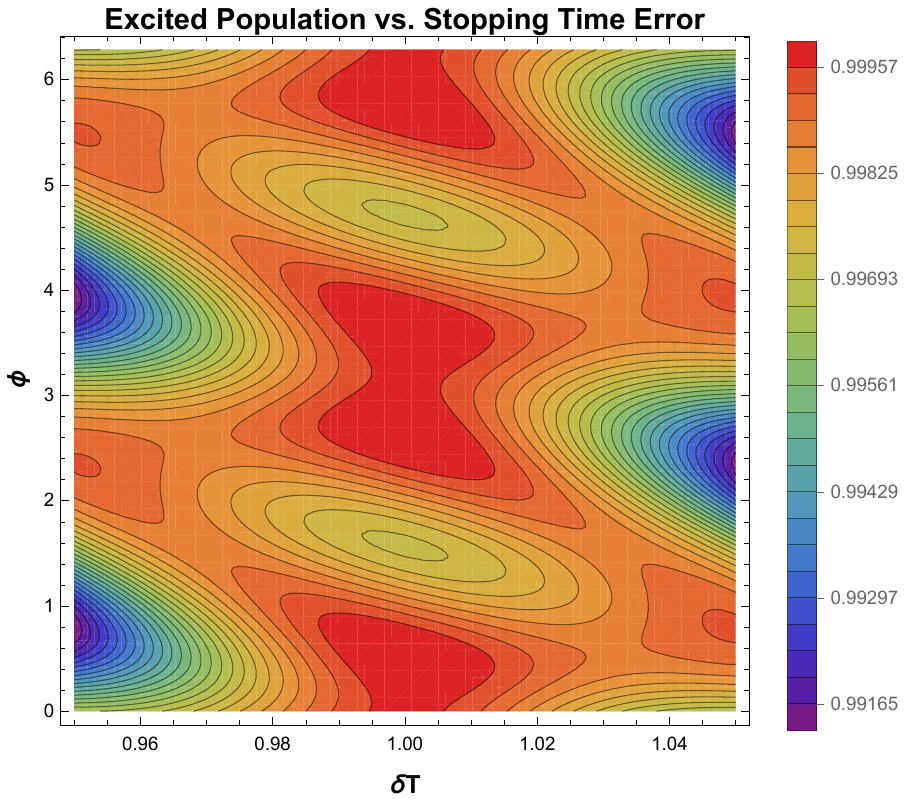}
    \caption{\textbf{Robustness with respect to small deviations in the amplitude $A$ and in the stopping time $T$.} The left panel shows that the protocol is robust for deviations in the amplitude $A$. The horizontal axis shows the deviation of $A$ in terms of fraction, where $1.0$ represents no error and $1.04$ represents the situation where the source has amplitude $1.04 A$ instead of $A$. Similarly, the right panel demonstrates that the protocol is robust for errors in the stopping time $T$. Both figures are drawn with $E=1$ and $A=0.05$.}
    \label{fig:amplitude_error_and_stopping_time_error}
\end{figure}

\medskip
\textbf{Robustness for a small deviation in the interaction strength $A$ and in the stopping time $T$.}
Our quantum DEH protocol is shown to be robust against small systematic errors in the source's amplitude $A$ and shift in the stopping time $T$. As we showed in the main text, the protocol must stop at $T= \pi/(2A)$, where $A$ defines the amplitude in the total Hamiltonian $H = -\frac{E}{2} Z + 2 A \cos (\omega t + \phi) X$. If the amplitude $A$ deviates slightly, say becomes $\delta A \times A $, but we still stop the protocol at $T=\pi/(2A)$, then the excited population will have an error. Similarly, deviation occurs in the excited state population when we stop at a time that is not exactly $T = \pi/ (2A)$, say at $\delta T \times T$. In Fig.~\ref{fig:amplitude_error_and_stopping_time_error}, we present the numerical analysis for the excited state population against small amplitude deviations and the stopping time, where $E=1$ and $A = 0.05$. We consider a deviation for the amplitude and the correct stopping time of no more than $5\%$, and the results suggest that the protocol can achieve the excited state with a probability of at least $0.99$.

\begin{figure}
    \centering
    \includegraphics[width=0.49\linewidth]{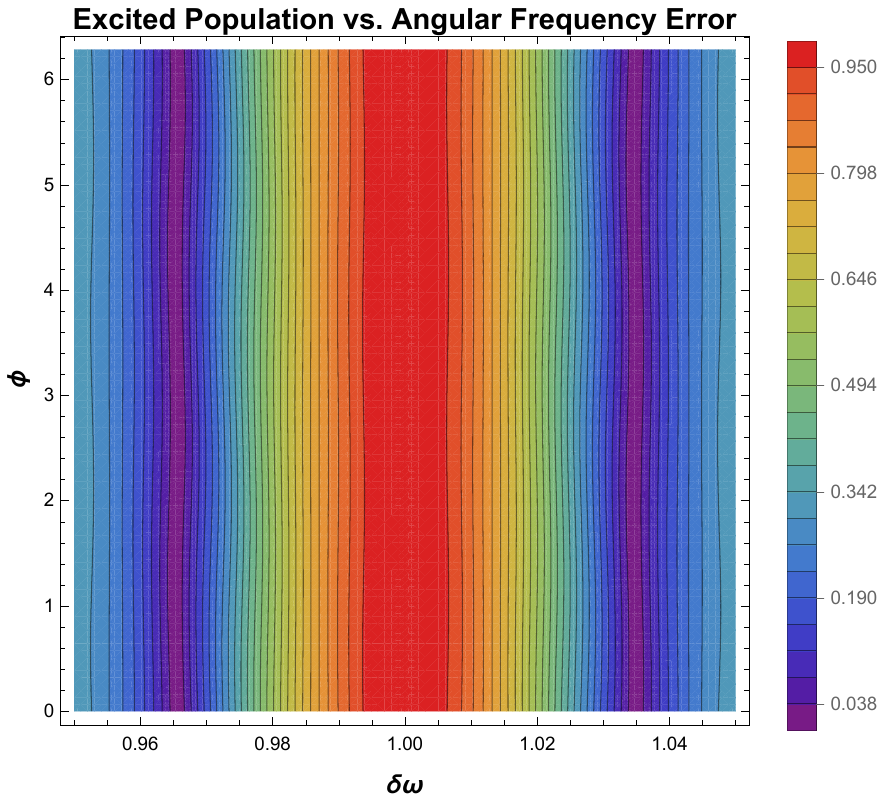}
    \includegraphics[width=0.49\linewidth]{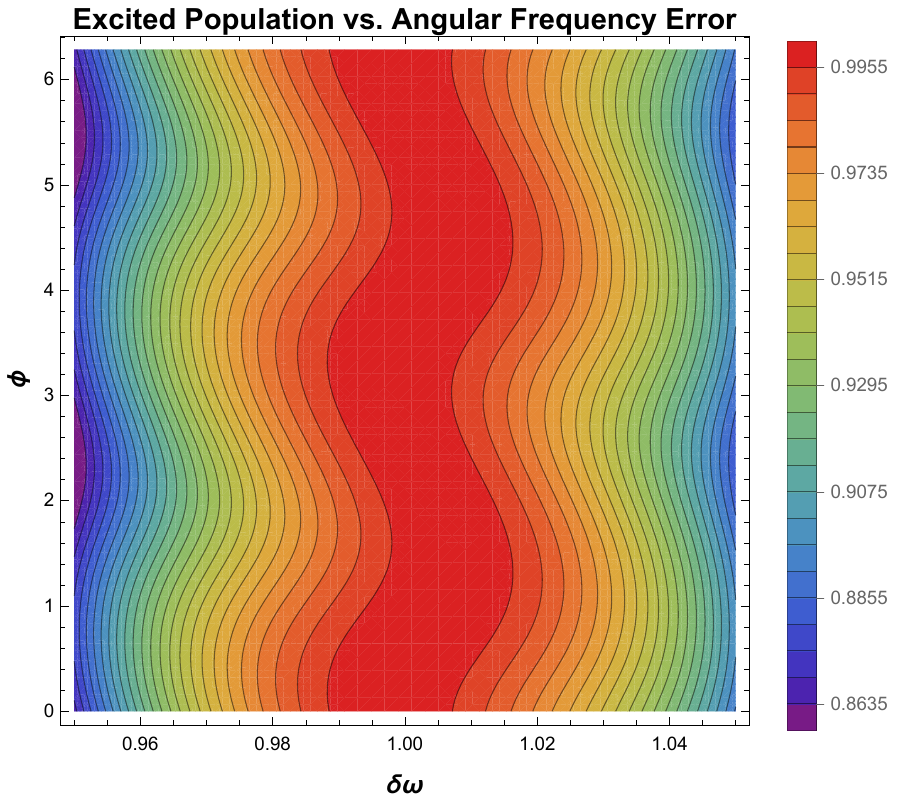}
    \caption{\textbf{Robustness with respect to the angular frequency shift.} Both panels plot the excited state population versus deviations in the angular frequency of the source $\omega$ with no larger than $5\%$, where $\delta \omega$ ranges from $95\%$ to $105\%$ and $\delta \omega = 1$ means there is no deviation. The left panel has $E=1$ and  $A=0.01$, and the right panel has $E=1$ and $A=0.05$.}
    \label{fig:angular_frequency_error}
\end{figure}

\medskip
\textbf{Robustness for a small deviation in the source's angular frequency $\omega$.} We also numerically analyze and demonstrate the robustness of our quantum DEH protocol against deviations in the angular frequency $\omega$ of the source. For the cases where $E=1$ and $A=0.01$ or $A=0.05$, results are shown in Fig.~\ref{fig:angular_frequency_error}.

\medskip
\textbf{Robustness of Quantum DEH Against Pulse Shaping}. 
During the Quantum DEH control interval $T = \frac{\pi}{2A}$, the amplitude of the fluctuating source $\cos(\phi + \omega t)$ might not instantly jump to $2A$ at the start point or drop to $0$ at the endpoint. 
Since an instantaneous signal switch-on/off is not feasible in actual physical setups, pulse shaping is necessary to ensure a smooth transition of the signal from zero to full amplitude.
This involves ramping the amplitude at both the beginning and end of the operation cycle.

\begin{figure}
    \centering
\includegraphics[width=0.7 \linewidth]{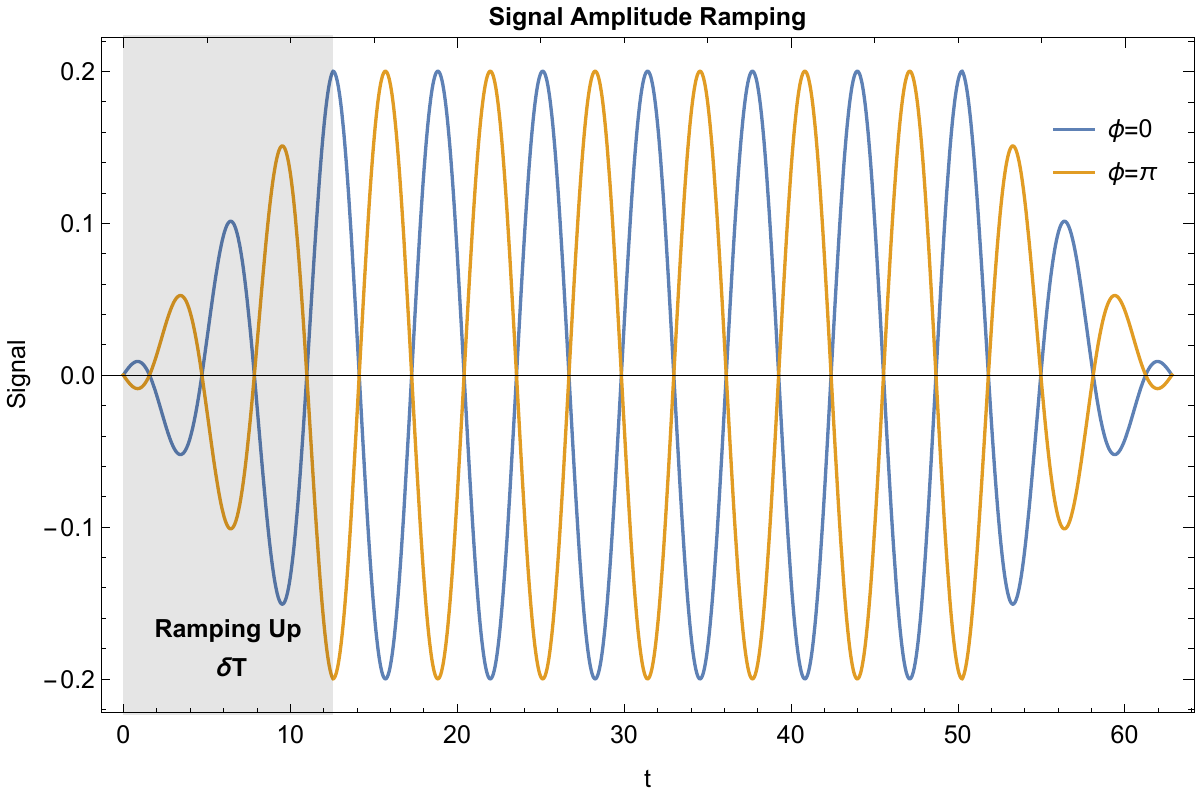}
\caption{\textbf{Ramping up the signal by increasing the amplitude gradually.} An amplitude modulation is added both in the beginning and the end of the signal to continuously switch on and off of the interaction. The ramping takes $\delta T$ to reach a stable sinusoidal signal. In this plot, $\delta T$ is set to be $20\%$ of the total time span $T$.}
\label{fig:RampingSignal}
\end{figure}

\begin{figure}
\centering
\includegraphics[width = 0.49 \linewidth]{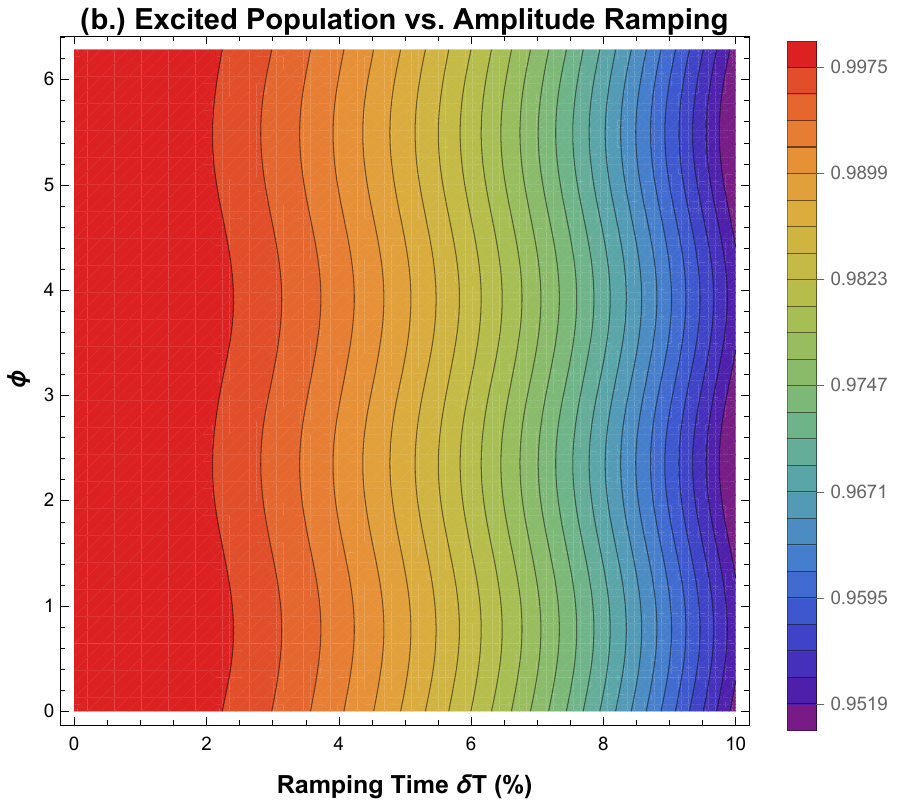}
\includegraphics[width = 0.49 \linewidth]{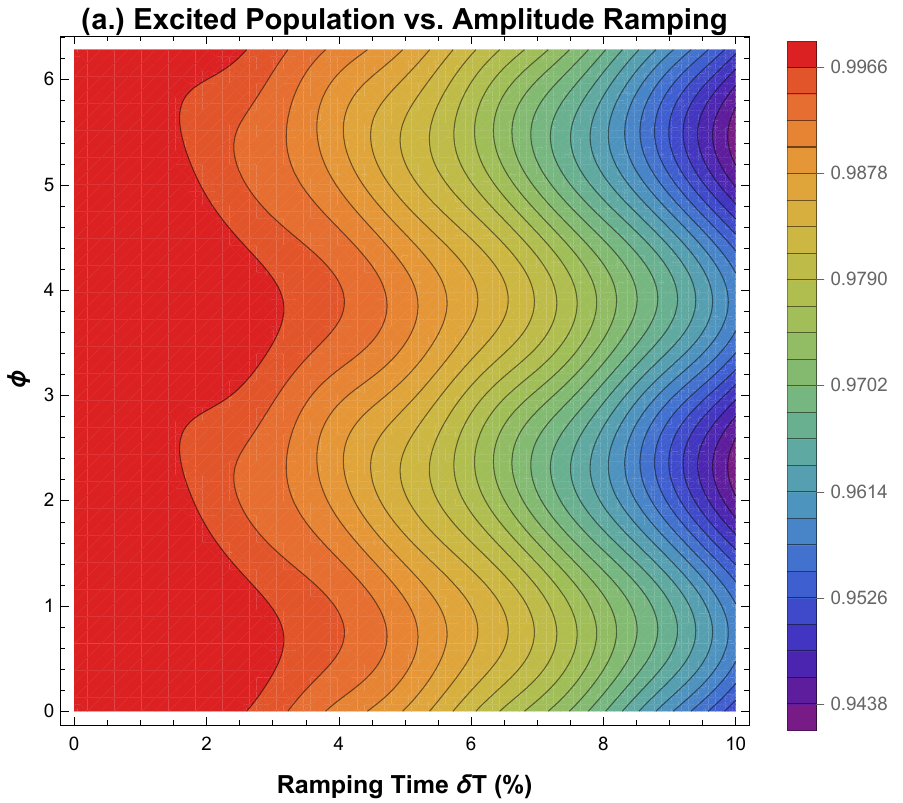}
\caption{\textbf{Excited population's robustness to pulse shaping.} The figure illustrates how the excited population varies with different ramping duration of amplitude modulation at the start and end of the protocol, expressed as a percentage of the total time period. Panel (a) shows results for amplitude \(A=0.05\), while panel (b) presents data for amplitude \(A=0.01\).}
\label{fig:Accuracy_pulseShaping_A0.01}
\end{figure}

As depicted in Fig ~\ref{fig:RampingSignal}, there is amplitude modulation at the beginning and the end of the operation. The signal starts at zero amplitude and gradually increases to full amplitude $2A$ over a period $\delta T \times T$, where $\delta T$ denotes the small fraction of the total operational time $T$. This modulation is a natural process and can be easily understood in a physical context:
Consider a two-level system interacting with a magnetic field generated by a sinusoidal current in a straight wire. If the time-dependent current is $I(t) = I_0 \cos (\omega t + \phi)$ with maximal amplitude $I_0$, then it will generate a radial magnetic field with strength $B = \frac{\mu_0 I(t)}{2\pi d} $ at radial distance $d$, where \(\mu_0\) is the permeability of vacuum free space. If the system starts far from the wire and moves closer within a time $\delta T \times T$, it will experience an increasing oscillatory signal, just as shown in Fig.~\ref{fig:RampingSignal}. Conversely, as the system moves away from the wire at the end of the protocol, the signal ramps down to zero. Therefore, such amplitude ramping can be implemented by moving the quantum system away and close to the wire, in which the sinusoidal fluctuating current with different phases flows through.
As Figure~\ref{fig:Accuracy_pulseShaping_A0.01} shows, our quantum DEH protocol demonstrates good robustness against pulse shaping.

\medskip
To conclude, the robustness against parameter deviation and the robustness of the pulse shaping make our quantum DEH protocol robust and suitable for practical applications.

\section{A family of external Hamiltonians for every $H_0$.}
\label{SM:VIII}

In this section, we distill the essential features of the qubit case with external potential $V=\cos(\omega t+\phi)$ case to generalize the set of potentials and quantum systems that can be used.

In the qubit $\cos(\omega t+\phi)$ case, an energy eigenstate of a bare Hamiltonian $H_0=\sum_{i=0,1} E_i\ket{i}\bra{i}$ is taken to another energy eigenstate. We can write the overall unitary as 
\begin{equation}
\label{eq:Ugen}
U=e^{i\theta}\ket{j}\bra{i}+e^{i\tilde{\theta}}\ket{i}\bra{j}+\sum_{k\neq i,j}\ket{k}\bra{k},
\end{equation}
for which one sees $U^{\dagger}U=\mathbb{I}$ as desired. Moreover, for any such $U$, 
\begin{equation}
\label{eq:UgenAction}
U\ket{i}\bra{i}U^{\dagger}=\ket{j}\bra{j},
\end{equation}
by inspection. 

For this overall unitary to increase the energy, we require $E_j>E_i$. A natural question is whether this can be achieved with a time-independent Hamiltonian. For the qubit case, one can visualize a family of unitaries corresponding to Pauli X conjugated by phase gates. Then there is a fixed bloch sphere rotation axis for each such unitary corresponding to the X direction and rotations around Z of the X axis. More generally, for any $U$ we can write 
\begin{equation}
    U=\exp(iA), 
\end{equation}
for some Hermitian $A=-i\ln U$. Suppose we want a time-independent potential $V$, then \begin{equation}
    A=H_0\tau+V\tau \, .
\end{equation}
So for any $(H_0,\tau)$, 
\begin{equation}
\label{eq:VU}
V_U=\frac{A-H_0\tau}{\tau}=\frac{-i\ln U-H_0\tau}{\tau}.
\end{equation}
Since there is a family of $U$'s of the form of Eq.\eqref{eq:Ugen}, there is a family of $V_U$'s which achieve the task. If the above is correct then for any $H_0$ we can find a family of time-independent $V_U$ such that the eigenstate of interest $\ket{i}$ is taken to the eigenstate $\ket{j}$ with certainty at time $\tau$. 

However, we also want the extra term to be off initially and at the end so that the change in energy is well-defined and equals $E_j-E_i$. We could artificially imagine that $V$ is switched on at t=0 and off at $t=\tau$, in which case $V$ is actually time-dependent. Thus it seems the above argument identifies a {\em time-dependent} family of potentials for any given $H_0,\tau$. 

For example, say we have a qutrit and 
\begin{equation}  H_0=E_1\ket{1}\bra{1}+E_2\ket{2}\bra{2}+E_3\ket{3}\bra{3}=(-1)\ket{1}\bra{1}+(0)\ket{2}\bra{2}+(1)\ket{3}\bra{3}.
\end{equation}
We demand
\begin{equation}
U=e^{i\theta}\ket{3}\bra{1}+e^{i\tilde{\theta}}\ket{1}\bra{3}+\ket{2}\bra{2},
\end{equation}
such that 
\begin{equation}
U\ket{1}\bra{1}U^{\dagger}=\ket{3}\bra{3}.
\end{equation}
Then, letting $\tau=1$ for simplicity
\begin{equation}
    V_U=-i\ln U-H_0.
\end{equation}

The above gives one general family of $V_U$ terms that work as a DEH, for a given $H_0$, $\tau$.

\section{Estimation of the power generated using the quantum DEH protocol.}
\label{SM:IX}

In this section, we estimate the power of the energy harvested using the quantum DEH protocol. 

Suppose we want to harvest energy from a linearly polarized electromagnetic plane wave with wavelength $\lambda$ and it travels along the $z$-direction, with the electric field polarized in the $x$-direction. The electric field $\mathbf{E}$ can be described by the following equation,
\begin{equation}
\mathbf{E}(z, t) = E_0 \cos\left( \omega t - \frac{2 \pi}{\lambda} z + \phi\right) \hat{\mathbf{x}} 
\end{equation}
where $E_0$ is the amplitude of the electric field, $\lambda =\frac{2 \pi c_0}{\omega}$ is the wavelength, $c_0$ is the speed of the light, $\phi$ is a phase constant, and $\hat{x}$ indicates that the field is polarized in the $x$-direction. For this plane wave, the average power per unit section area is
\begin{equation}
    P_{in} = \langle S \rangle = \sqrt{\frac{\epsilon_0}{4 \mu_0}} E_0^2 \, , \label{eq:Power_input}
\end{equation}
where $S$ is the magnitude of the Poynting vector  $\mathbf{S=E \times B}$, and $\mu_0$ and $\epsilon_0$ are the vacuum permeability and vacuum permittivity respectively.
When we fix our energy harvester at the $z=0$ plane, then it will experience a sinusoidal fluctuating $\mathbf{E}$ field,
\begin{equation}
    \mathbf{E} (t) = E_0 \cos (\omega t + \phi ) \hat{\mathbf{x}} \, .
\end{equation}

Now we estimate the power of the quantum energy harvester.
The Hamiltonian for a quantum dipole as a two-level system experiencing the sinusoidal electric field along the $\hat{\mathbf{x}}$ direction at resonance is
\begin{equation}
    H = -\frac{\hbar \omega}{2} Z + E_0 d \cos (\omega t + \phi ) X \, ,
\end{equation}
where $d$ is the magnitude of the effective dipole moment of the two-level system, $X$ and $Z$ are the Pauli matrices and we assume the ground state is $\ket{0}$. According to our result in the main text, this quantum harvester will collect $\hbar \omega$ amount of energy after evolving for $T = \frac{\pi \hbar}{ E_0 d}$ amount of time. So the power of one electric dipole as the quantum harvester is 
\begin{equation}
    P_{\text{quan, per dipole}} = \frac{ E_0 d \omega}{ \pi} \, .
\end{equation}
Now suppose that there are $n$ quantum dipoles per unit area on the $z=0$ plane that are used to extract energy, then the power collected per unit area is
\begin{equation}
    P_{\text{quan, per area}} = \frac{E_0  d  n \omega }{\pi } \, .
\end{equation}

Now, we apply the concrete model to get an estimation of the value of the power. Because of its potential for high-density integration, we consider the Rabi oscillation of artificial dipole by quantum dots~\cite{stievater2001rabi}, which are implemented by GaAs semiconductor quantum dots of $10$nm (nanometer) scale. As a rough estimation, we assume that the quantum dots are separated from each other with 20nm distance and we assume that there is no mutual influence, which leads to a density of the harvester $n = \frac{1}{400 \mathrm{nm}^2}$. According to the experiment in Ref~\cite{stievater2001rabi}, the energy gap $\hbar \omega $ is estimated to be 1 meV, and the dipole moment $d$ is estimated to be 75 Debye. Assuming the intensity of the electromagnetic wave is $I = 1000 W/m^2$, which can be expressed using the amplitude of the electric field,
\begin{equation}
    I = \frac{1}{2} c_0 \epsilon_0 E_0^2 \, ,
\end{equation}
where $\epsilon_0$ is the vacuum permittivity. Putting the numbers in, we have
\begin{equation}
    P_{\text{quan, per area}} = 41.8 W/m^2 \, ,
\end{equation}
and the power per dipole as
\begin{equation}
    P_{\text{quan, per dipole}} = 1.671 \times 10^{-14} W \, .
\end{equation}

\end{document}